\documentclass[times,12pt]{article}

\usepackage{dsfont}
\usepackage{units}
\usepackage{uri}
\usepackage[margin=2.5cm]{geometry}
\usepackage{natbib}
\usepackage{amstext, amsmath}
\usepackage{graphicx}
\usepackage{booktabs}
\usepackage[font={small,it}]{caption}

\begin{document}

\title{Rapid adjustment and post-processing of temperature forecast trajectories}

\author{N.~Schuhen \and  T.~L.~Thorarinsdottir \and A.~Lenkoski }

\date{Norwegian Computing Center, Oslo, Norway}

\maketitle

\begin{abstract}
  \noindent
Modern weather forecasts are commonly issued as consistent multi-day forecast trajectories with a time resolution of 1--3 hours. Prior to issuing, statistical post-processing is routinely used to correct systematic errors and misrepresentations of the forecast uncertainty. However, once the forecast has been issued, it is rarely updated before it is replaced in the next forecast cycle of the numerical weather prediction (NWP) model. This paper shows that the error correlation structure within the forecast trajectory can be utilized to substantially improve the forecast between the NWP forecast cycles by applying additional post-processing steps each time new observations become available. The proposed rapid adjustment is applied to temperature forecast trajectories from the UK Met Office's convective-scale ensemble MOGREPS-UK. MOGREPS-UK is run four times daily and produces hourly forecasts for up to 36 hours ahead. Our results indicate that the rapidly adjusted forecast from the previous NWP forecast cycle outperforms the new forecast for the first few hours of the next cycle, or until the new forecast itself can be rapidly adjusted, suggesting a new strategy for updating the forecast cycle. 
\end{abstract}

\section{Introduction}

Weather forecasts resulting from numerical weather prediction (NWP) models are traditionally post-processed using statistical approaches in order to correct potential systematic biases in the forecasts \citep{GlahnLowry1972}. Roughly 15 years ago, the first papers on statistical post-processing methods yielding full predictive distributions--correcting both systematic biases and assessments of forecast uncertainty--appeared in the literature \citep{Gneiting&2005, Raftery&2005}. Since, approaches of that type have become increasingly more common in both the literature and operational forecasting for NWP forecasts and forecast ensembles \citep{Vannitsem&2018}. Originally, the methods applied to marginal predictive distributions of individual weather variables at individual locations \citep{Gneiting&2005, Raftery&2005}. More recent work has produced consistent probabilistic predictions for temporal trajectories \citep{Hemri&2015}, spatial forecast fields \citep{Berrocal&2008, Feldmann&2015} and multiple variables \citep{Schuhen&2012, Moeller&2013, Sloughter&2013}. See \cite{Vannitsem&2018} for a recent overview of statistical post-processing methods for ensemble forecasts.

\begin{figure*}
  \centering
  \includegraphics[width=0.7\textwidth]{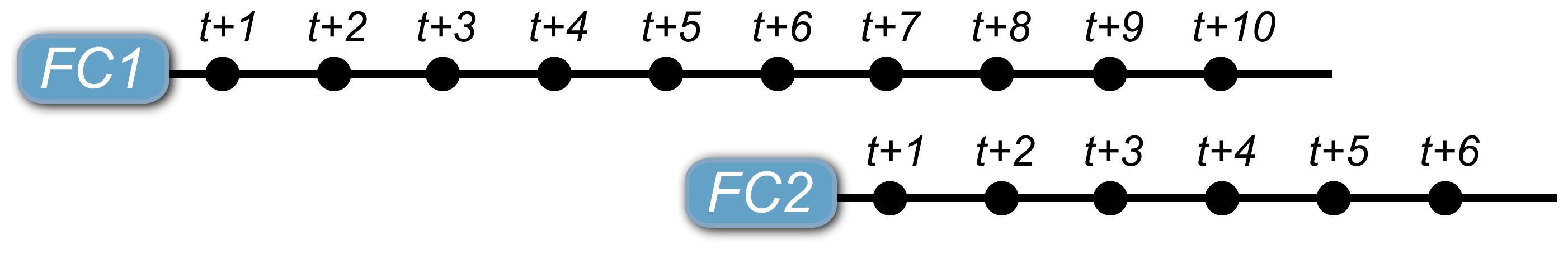}
  \caption{Diagram of a typical forecast cycle for hourly forecasts issued every six hours. The MOGREPS-UK version used in this paper is configured in this way.}\label{fig:fig01}
\end{figure*}

The aim of probabilistic forecasting is to ``maximize the sharpness of the predictive distribution subject to calibration'' \citep{Gneiting&2007}. Here, calibration, or reliability, refers to the statistical consistency between the forecast and the observation; a forecast is (probabilistically) calibrated if events predicted to have probability $p$ are realized with the same relative frequency in the observations. A calibrated forecast should then provide as much information regarding future weather as possible; the smaller the forecast uncertainty, or the higher the sharpness of the predictive distribution, the more information regarding future weather is contained in the forecast. In practice, the NWP model outputs a forecast trajectory for multiple lead times. As soon as the model output is available, the forecasts of the entire trajectory are post-processed using the most recent available pairs of previous forecasts and verifying observations to obtain calibrated and sharp forecasts for all lead times. A new, post-processed forecast is then issued for all future time points corresponding to the lead times of the original NWP forecast. An example of such a setting is shown in Figure~\ref{fig:fig01} for an hourly forecast where a new forecast is issued every six hours.     

In the standard setting demonstrated in Figure~\ref{fig:fig01}, the published forecast is not updated until it is replaced in the next forecast cycle of the NWP model. However, new information in the form of new observations becomes available every hour. In the current paper, we propose an approach for rapid adjustment of forecast trajectories (RAFT), where, in addition to standard post-processing, we regularly update the forecast every time a new piece of information becomes available by utilizing the positive error correlation of the forecast errors within an NWP forecast trajectory. The idea behind RAFT is related to that of data assimilation, see e.g. \cite{MitchellHoutekamer2000} who develop a method to account for model error in the context of an ensemble Kalman filter technique. Here, our main priority is computational efficiency to minimize the time needed for each adjustment. We thus propose an efficient adjustment approach that is adapted to each forecast cycle, hour and lead time separately. In a case study, we apply the method to hourly temperature forecasts from the MOGREPS-UK ensemble from the UK Met Office whose schedule follows the forecast cycle shown in Figure~\ref{fig:fig01}. 

The remainder of the paper is organized as follows. In the next Section~\ref{sec:sec02}, we introduce the MOGREPS-UK forecast ensemble and the corresponding observations, and review the classical ensemble model output statistics (EMOS) post-processing method as well as the validation metrics used in our study. We further show the skill of the post-processed EMOS forecasts. In Section~\ref{sec:sec03}, we introduce our proposed method for rapid adjustment of forecast trajectories (RAFT). Results at Heathrow Airport as well as those over the entire study region are presented in the following Section~\ref{sec:sec04}. Finally, the paper concludes with a summary and discussion in Section~\ref{sec:sec05}. 

\section{Data and conventional post-processing}\label{sec:sec02}

\subsection{MOGREPS-UK}

\begin{figure}
  \centering
  \includegraphics[width=0.5\textwidth]{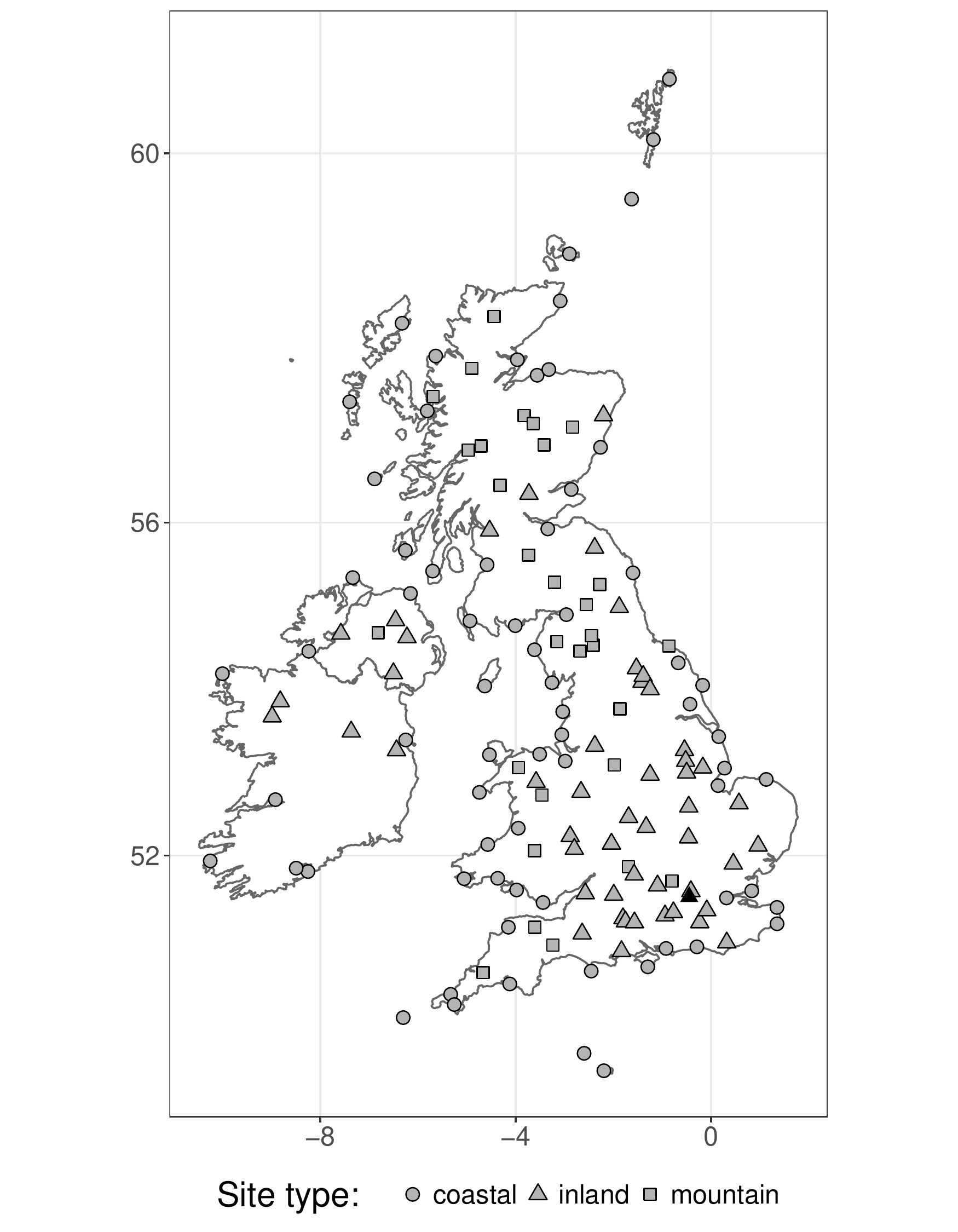}
  \caption{Map of the 150 observation locations in the UK and the Republic of Ireland used in this study. The sites are divided into three categories: coastal, inland and mountain sites. The black triangle marks Heathrow Airport.}\label{fig:fig02}
\end{figure}

Our data set consists of surface temperature forecasts and observations for 150 locations in the UK and the Republic of Ireland. The forecasts are provided by the UK Met Office's convective-scale ensemble MOGREPS-UK \citep{Hagelin&2017}, which has been running operationally since July 2012. The data set covers a period of 30 months between January 2014 and June 2016, during which the ensemble had a horizontal resolution of \unit{2.2}{km} and produced hourly forecasts for up to 36 hours. MOGREPS-UK is run 4 times daily, at 0300, 0900, 1500 and 2100 UTC. The initial and boundary conditions were originally provided by the global MOGREPS-G ensemble, but since March 2016 the high-resolution UKV analysis is used to initialize MOGREPS-UK. The ensemble consists of one control forecast and eleven perturbed members, which are generated from a combination of analysis increments and perturbations from the global ensemble. For the purpose of this paper, we treat them as twelve exchangeable ensemble members.

In this study, we consider site-specific data only, interpolated by the Met Office from model grid to observation locations. During this process, forecasts are corrected for local effects and the height differences between station and model orography. The observations are extracted from SYNOP messages at the 150 locations in Figure~\ref{fig:fig02} and Met Office quality controls have been applied. We separate the data into a training set (January to December 2014) with 7,018,719 forecast-observation pairs and a test set (January 2015 to June 2016) with 11,320,762 forecast-observation pairs. Although there have been several operational changes to the MOGREPS-UK model during these periods, we treat the data set as homogeneous over the entire study period.

\subsection{Ensemble model output statistics}\label{sec:sec2_2}

For all their benefits, weather forecast ensembles are usually too confident and produce underdispersed forecasts \citep{Hamill2001}. This means that the ensemble spread does not cover all sources of uncertainty in a given weather situation and is therefore on average too narrow. Like all weather prediction models, ensembles are also subject to a deterministic bias, depending on the model's skill in varying weather situations. To correct for the bias and the underdispersion, we first apply statistical post-processing to the raw ensemble forecasts before using the new RAFT error correction method. Ensemble model output statistics \citep[EMOS;][]{Gneiting&2005}, sometimes called non-homogeneous Gaussian regression, has successfully been applied to multiple forecast models \citep[e.g.\ ][]{Feldmann&2015,ScheuererBuermann2014,Kann&2009} and is a suitable method to calibrate MOGREPS-UK forecasts.

We denote a future temperature observation for a specific location and time by $Y$ and the corresponding ensemble forecast members by $X_1,\ldots,X_{12}$. The EMOS predictive distribution of $Y$ conditional on $X_1,\ldots,X_{12}$ is then defined as a Gaussian distribution:
\begin{equation}
Y \mid X_1,\ldots,X_{12} \sim \mathcal{N} \left( \mu, \sigma^2 \right)
\end{equation}
The moments of this distribution are modeled using the ensemble forecast's statistics; the predictive mean
\begin{equation} \label{eq:predmean}
\mu = a + b^2 \cdot \bar{X}
\end{equation}
is a linear function of the ensemble mean $\bar{X}=\frac{1}{m} \sum_{i=1}^{m}X_i$ and the predictive variance 
\begin{equation} \label{eq:predvar}
\sigma^2 = c^2 + d^2 \cdot S^2
\end{equation}
an affine function of the ensemble variance $S^2=\frac{1}{k}\sum_{i=1}^{m}\left( X_i - \bar{X} \right)^2$. Here, $m=12$ is the number of ensemble members and the coefficients $a$, $b$, $c$ and $d$ are real numbers. For estimating $a$, $b$, $c$ and $d$, we use minimum score estimation \citep{Dawid&2016} and optimize the continuous ranked probability score \citep[CRPS;][]{MathesonWinkler1976,GneitingRaftery2007} based on training data. The parameters in Equation~\ref{eq:predvar} are squared to ensure that the predictive variance is non-negative. In Equation~\ref{eq:predmean}, $b$ is constrained in the same way, making it easier to interpret.

All runs of the NWP model and all forecast lead times are calibrated separately using a rolling training period of 40 days. This means that for each run and each lead time, we collect all forecast-observation pairs from the last 40 days, where the forecasts were initialized at the same time of day and are valid for the same lead time. This data comprises the basis for the estimation of the EMOS coefficients. The current ensemble forecasts are then plugged into Equations~\ref{eq:predmean} and \ref{eq:predvar} to obtain the full EMOS predictive distribution $\mathcal{N} \left( \mu,\sigma^2 \right)$. We follow the local EMOS approach, in that all stations are treated on an individual basis. This accounts for local effects and turns out to produce much better results than a regional approach, where data from different sites is pooled together. In order to have a full set of training data for the first model runs in 2014, some dates from the end of 2013 are used.

\subsection{Verification methods and EMOS forecast skill}
To evaluate the effectiveness of the EMOS method, we compare the predictive skill of the post-processed forecasts to the raw MOGREPS-UK ensemble. The tools used here, as well as for evaluating the RAFT forecasts in Section~\ref{sec:sec04}, are the root-mean-square error (RMSE), the CRPS and the rank and probability integral transform (PIT) histograms. Both the RMSE and the CRPS are proper scoring rules \citep{GneitingRaftery2007}, they measure the skill of a forecast by assigning a numerical penalty depending on how well the forecasts match the observations. It is essential that they are proper, as this guarantees that the best forecast model will receive the best score and prohibits hedging. While the RMSE assesses the deterministic forecast accuracy of the mean of a predictive probability distribution $F$, the CRPS evaluates the probabilistic skill of the whole distribution--which can also be represented by a discrete ensemble. The RMSE is defined as the square root of the average squared distance between the mean forecasts and the observations $y$:
\begin{equation}
\text{RMSE} \left( F, y \right) = \sqrt{\frac{1}{n}\sum_{i=1}^{n}(\text{mean} \left( F \right) - y)^2},
\end{equation}
where $n$ is the number of data points or forecast cases.

In its general form, the CRPS can be expressed as the squared area between a forecast cumulative distribution function (CDF) $F$ and the empirical CDF of the observation $y$ \citep{ThorarinsdottirSchuhen2018}:
\begin{equation}
\textup{CRPS} \left( F,y \right) = \int_{-\infty}^{+\infty} \left( F\left( x \right) - \mathds{1} \left\{ y \leq x \right\} \right) ^2 \, \text{d} x
\end{equation}
Here, we use the closed form for a Gaussian distribution (Equation~\ref{eq:CRPS_EMOS}) to evaluate the EMOS forecasts and an approximation for the MOGREPS-UK forecasts, where the distribution is given by an ensemble (Equation~\ref{eq:CRPS_ENS}):
\begin{align}
\text{CRPS}_\text{EMOS} \left( \mathcal{N} \left( \mu, \sigma^2 \right) ,y \right) & = \sigma \left\{ \frac{y-\mu}{\sigma} \left[ 2\Phi \left( \frac{y-\mu}{\sigma} \right) -1 \right] + 2\phi \left( \frac{y-\mu}{\sigma} \right) - \frac{1}{\sqrt{\pi}} \right\} \label{eq:CRPS_EMOS}\\
\text{CRPS}_\text{ENS} \left( X_1, \ldots, X_m; y \right) &= \frac{1}{m} \sum_{i=1}^{m} \left| X_i-y \right| - \frac{1}{2m^2} \sum_{i=1}^{m} \sum_{j=1}^{m} \left| X_i - X_j \right| \label{eq:CRPS_ENS}
\end{align}
The functions $\Phi\left( \cdot \right)$ and $\phi \left( \cdot \right)$ in Equation~\ref{eq:CRPS_EMOS} indicate the CDF and the probability density function (PDF) of a standard Gaussian distribution, respectively.

\begin{table}
  \caption{Continuous ranked probability score (CRPS) and root-mean-square error (RMSE) averaged over all sites and forecast runs, for different lead time ranges. The margin of error based on a 95\% bootstrap interval is less than 0.002.}\label{tab:tab01}
  \centering
  \begin{tabular}{l | c c c | c c c}
    \toprule
     & \multicolumn{3}{c} {CRPS} & \multicolumn{3}{c}{RMSE} \\
    Lead times & 1--12 h & 13--24 h & 25--36 h & 1--12 h & 13--24 h & 25--36 h\\
    \midrule
    Raw ensemble & 0.718 & 0.741 & 0.792 & 1.205 & 1.254 & 1.343\\
    EMOS & 0.555 & 0.596 & 0.636 & 1.054 & 1.131 & 1.204\\
    \bottomrule
  \end{tabular}
\end{table}

Table~\ref{tab:tab01} summarizes both scores for the EMOS and the raw MOGREPS-UK forecasts. We divide the forecast lead times into three categories, early (1 to 12 hours), mid-range (13 to 24 hours) and later lead times (25 to 36 hours) and average the scores over each of the categories. As can be expected, the scores deteriorate with increasing lead time, for both EMOS and raw ensemble forecasts. By applying the EMOS post-processing technique, the probabilistic forecast skill is improved by around 20\% and the deterministic skill of the mean forecast by around 10\%.

With the addition of the new RAFT technique, applied directly to the mean of the EMOS forecast distribution, we seek to increase the deterministic skill even further, which in turn also leads to a reduction in the CRPS. However, we need to make sure that the changes do not affect the calibration of the probabilistic forecasts. To this end, we use the PIT histogram to check the level of calibration \citep{ThorarinsdottirSchuhen2018}. For a perfectly calibrated forecast, the PIT values, computed by evaluating the forecast CDFs at the observations, should form a flat histogram. The equivalent method for discrete ensemble forecasts is the verification rank histogram \citep{Anderson1996,HamillColucci1997,Talagrand&1997}, which measures the distribution of the observation rank in the set of ensemble forecasts. Both histograms are interpreted in the same way. 

\begin{figure*}
  \centering
  \includegraphics[width=\textwidth]{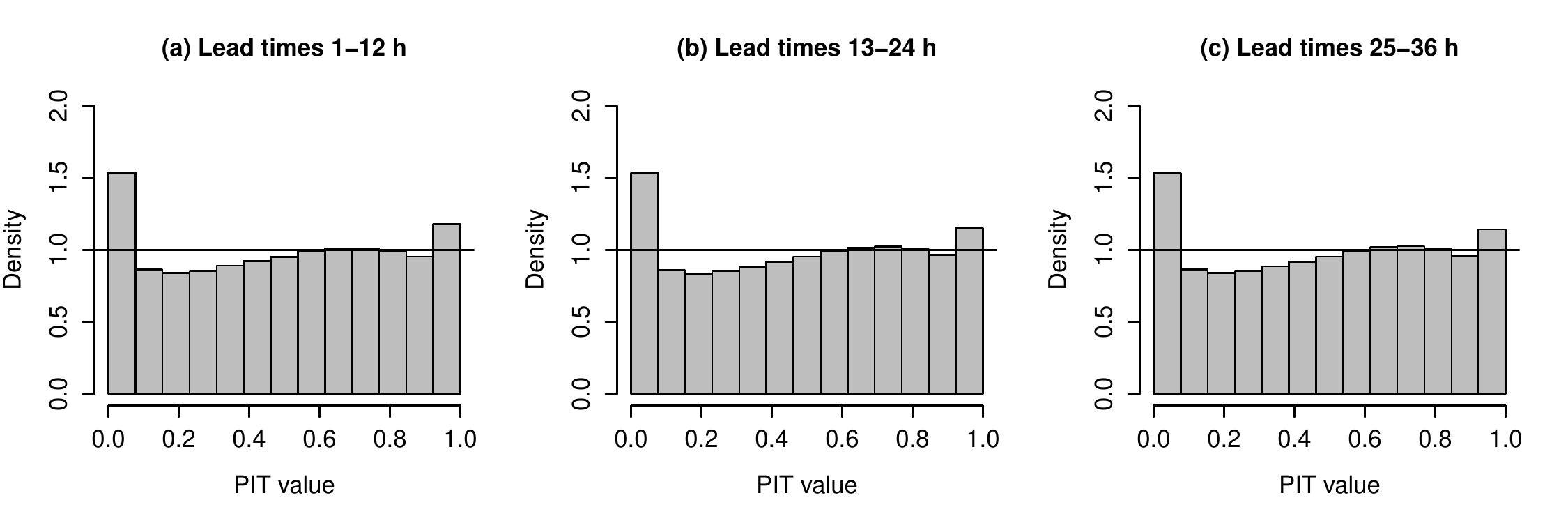}
  \caption{Probability integral transform (PIT) histograms of the EMOS post-processed forecasts, indicating the degree of calibration. Forecast cases are aggregated over all sites and forecast runs in the test set for (a) early lead times, (b) mid-range lead times and (c) later lead times} \label{fig:fig03}
\end{figure*}

In Figure~\ref{fig:fig03}, the PIT histograms for the EMOS forecasts are shown. Overall, they seem reasonably flat, however it seems that small miscalibrations remain in that there are, in particular, too many observations that land in the lower tail of the predictive distribution. There is almost no difference in the degree of calibration for the different lead times categories. These results indicate that a major jump in forecast skill can be achieved by applying EMOS to the raw ensemble. In a next step, the forecast trajectories provided by the EMOS mean are successively updated using the RAFT technique. Therefore, EMOS forms a baseline against which all further error reduction is measured. 

\section{Rapid adjustment of forecast trajectories}\label{sec:sec03}

The goal of the new RAFT method is to adjust and improve forecast trajectories over time by using the part of the trajectory that has already verified, in conjunction with the matching observations. First we need to establish the relationship between forecast errors at different lead times. The forecast error $e_{t,l}$ is here defined as the distance of the EMOS mean forecast $\mu_{t,l}$ to the observation $y_{t+l}$, where the forecast is initialized at time $t$ and valid at lead time $l$:
\begin{equation}\label{eq:error_obs}
e_{t,l} = y_{t+l} - \mu_{t,l}
\end{equation}

\begin{figure*}
 \centering
 \includegraphics[width=1\textwidth]{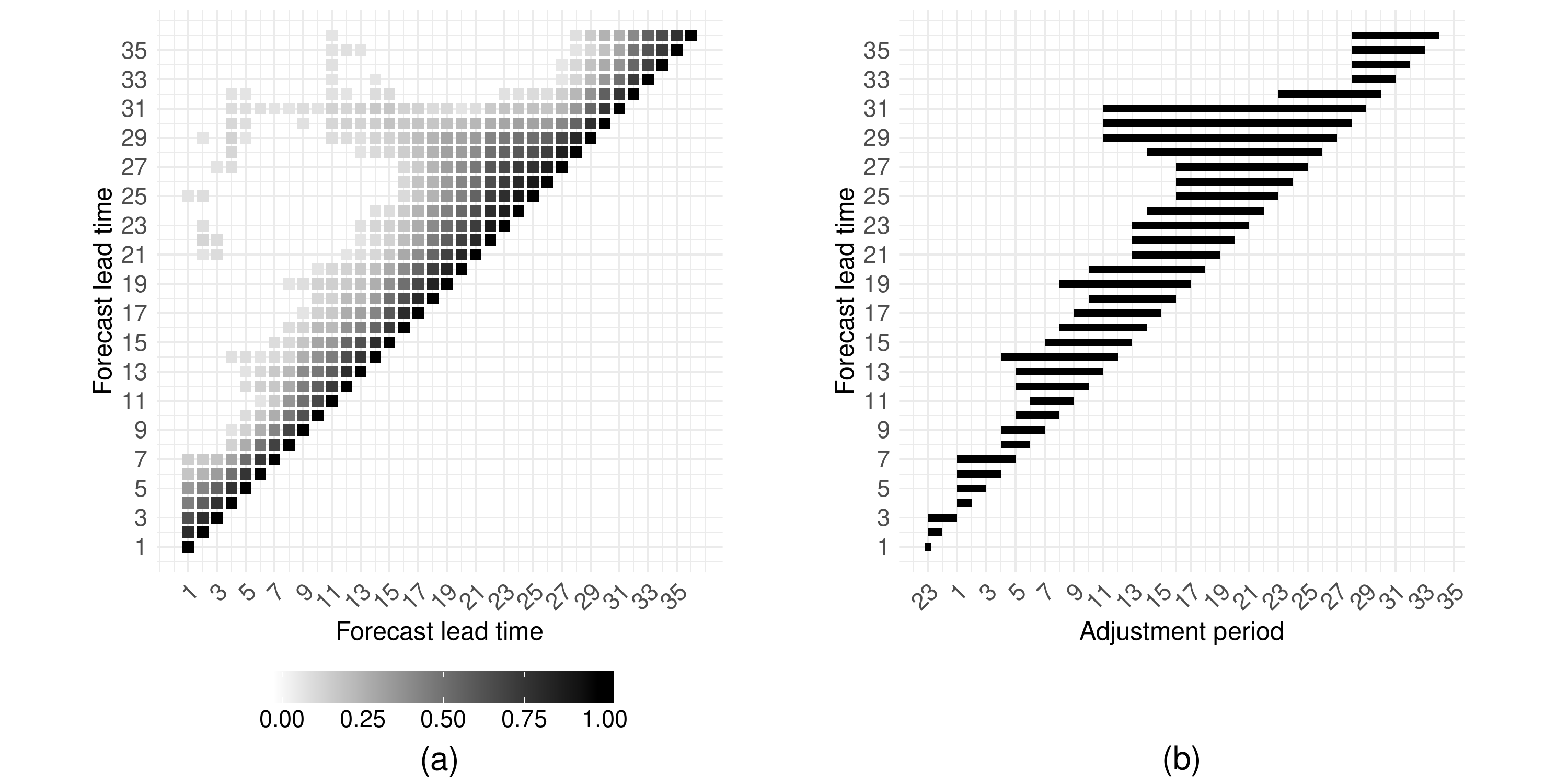}  
 \caption{(a) Empirical correlation coefficient of the forecast error for every lead time combination of the 0300 UTC model run at Heathrow Airport during the training period from January to December 2014. Correlations are only plotted if they are significant at the 90\% level. (b) RAFT adjustment period for each forecast lead time for the 0300 UTC model run at Heathrow Airport. The periods refer to the time points, where the observations used to adjust the future forecast are recorded.}
 \label{fig:fig04}
\end{figure*}

Figure~\ref{fig:fig04}a shows the Pearson correlation coefficient matrix of the forecast errors at Heathrow Airport (marked with a black triangle in Figure~\ref{fig:fig02}) for the 0300 UTC model run. To create the plot, the error correlations for all possible pairs of lead times were computed over the training set, as well as the corresponding p-values. Only statistically significant correlations at the 90\% level are shown. The correlation between lead times 1 and 36 is slightly negative and significant, but is left out for clarity and ultimately has no relevance for this study. 

In all instances, there is a positive correlation between the errors at a certain lead time and its immediate neighbors. This means that the errors at two lead times, if close enough, are so strongly connected that we can make inference about the forecast skill at a future lead time by observing the error at the earlier lead time. Formally, there is a period preceding each forecast $\mu_{t,l}$, during which the recently measured forecast error $e_{t,l^*}$, with $l^* < l$, provides useful information for a forecast adjustment at time $t+l$ and thus can reduce the subsequent error $e_{t,l}$. 

The size of these temporal neighborhoods varies greatly with the time of day. At lead times 8 to 11, corresponding to midday, the relationship between the forecasts is weakest with only 4 to 5 hours of significant correlation, while the largest predictability of 15 to even 27 hours can be found at lead times 28 to 31, in the early morning. In the MOGREPS-UK setting, this makes the RAFT method work on a rather short time scale, adjusting forecasts sometimes at only a couple of hours in advance. However, RAFT adapts to the scale and context of the application, e.g. for daily weather forecasts, the potential time range of adjustment increases to a few days.

Based on the correlation structure in Figure~\ref{fig:fig04}a, we can now define the RAFT model, establishing the relationship between forecast errors at two different lead times by linear regression. The estimated future error $\hat{e}_{l}$ at lead time $l=1,\dots,36$ is written as a linear function of the observed error at earlier lead times $l^*$:
\begin{equation} \label{eq:raft_regression}
\hat{e}_{l} = \hat{\alpha} + \hat{\beta} \cdot e_{l^*} + \varepsilon
\end{equation}
The error term $\varepsilon$ is normally distributed with mean zero and both coefficient estimates $\hat{\alpha}$ and $\hat{\beta}$ are determined by the least squares approach based on the training data set. All lead time combinations, sites and the four NWP model initialization times are treated separately. We omit the index $t$ for the model run from Equation~\ref{eq:raft_regression} for simplicity. Regression equations using multiple lead times as predictors were also investigated, but did not yield any improvement, as the newest observation always contains the most useful information.

Not all of the possible lead time combinations produce valid and useful results. As seen in Figure~\ref{fig:fig04}a, the correlation between lead times, and therefore predictability, is irregular and depends on various factors. Consequently, we define for each lead time $l$ an adjustment period of length $p$, consisting of the preceding lead times for which there is a strong enough correlation to affect the forecast skill. Starting at $l-p+1$, the forecast for lead time $l$ is repeatedly adjusted in hourly steps, each time using the most recent available forecast error information. Here, we allow for a processing time of one hour after an observation has been recorded, which means that the final adjustment for a forecast valid at lead time $l$ is made at $l-1$, using the error at $l-2$.

To establish the length of the adjustment period for each location and lead time, we use the following algorithm:
\begin{enumerate}
  \item Run the linear regression in Equation \ref{eq:raft_regression} using all lead times $l^* \in [l-23,l-2]$ as predictors. For negative lead times, add 24 hours, so that lead time 23 is followed by lead time 0, 1, etc.
  \item
  \begin{enumerate}
    \item Working backwards, find the first instance of $l^*$ in $[l-11;l-2]$ where the regression coefficient $\hat{\beta}$ is not significantly different from zero at the 90\% level. If a result can be found, we denote it by $l_p$.
    \item If such an $l_p$ can not be found, find the first instance of $l^*$ in $[l-19;l-12]$ where $\hat{\beta}$ is not significantly different from zero at the 95\% level. If a result can be found, we denote it by $l_p$.
    \item If such an $l_p$ can not be found, find the first instance of $l^*$ in $[l-23;l-20]$ where $\hat{\beta}$ is not significantly different from zero at the 99\% level.
  \end{enumerate}
  \item Set $p=l-l_p$. If no value for $l_p$ is found after Step 2, then $p$ is the average of the adjustment period lengths of the neighboring lead times $l-1$ and $l+1$. In case this does not produce a valid number, $p$ is set to $22$, the maximum possible length for the adjustment period.
\end{enumerate}
Establishing that $\hat{\beta}$ should be different from zero ensures that any adjustment made is not random, but based on genuine additional error information. This somewhat arbitrary algorithm was designed so that it works well for a multitude of sites in our data set with very different correlation patterns. It can be replaced by any other method for identifying a suitable adjustment period. Figure~\ref{fig:fig04}b shows the adjustment periods for the 0300 UTC run at Heathrow Airport produced by the algorithm above. It is clear that there is a strong connection between the correlation pattern in Figure~\ref{fig:fig04}a and the adjustment period length, in that large $p$ correspond to longer periods of predictability. 

The adjustment period refers to the time points when the observations used in the adjustment are recorded, and not the time points when the adjustments are carried out. As we allow an extra hour for the processing of the observations, the actual correction is made one hour after the observation time, starting at $l-p+1$. For example, we see from Figure~\ref{fig:fig04}b that the ideal length of the adjustment period for lead time $l=25$ here is $p=9$. This means that the first correction to a forecast valid at $t+25$ is made at $t+l-p+1=t+17$ using the observation collected one hour earlier, at $t+16$. From there on, an adjustment takes place every hour, each time using the newest error information available at that moment, until the time $t+24$, where we adjust the forecast for a final time based on the error measured at $t+23$. Clearly this last observation gives us the most accurate information about the expected forecast error, as it is closest in time to the forecast. This means that we get the most gain in forecast skill if RAFT is applied in the very short term.

Obviously, there is a gap during the first two hours of the forecast trajectory, where no forecast data from the current run is available to adjust the forecasts at $t+1$ and $t+2$. In this case, we instead use forecasts from the run that was initialized 24 hours earlier which are valid at the same time as the missing forecasts. Of course this does not lead to the same kind of improvement in forecast skill, as the current forecast run might exhibit a very different error characteristic than the one from 24 hours ago.

To obtain the size and direction of the forecast adjustment for a certain forecast run $t$ and lead time $l$, we first calculate the observed error $e_{t,l-k}$ at lead time $l-k$ according to Equation~\ref{eq:error_obs}, where $k \leq p$ and the time $l-k$ thus lies within the adjustment period. Then we plug the observed error into the regression equation for the predicted error $\hat{e}_{t,l}$ at the future time point $t+l$: 
\begin{equation}
\hat{e}_{t,l} = \hat{\alpha} + \hat{\beta} \cdot e_{t,l-k}.
\end{equation}
The regression coefficient estimates $\hat{\alpha}$ and $\hat{\beta}$ are unique for each lead time combination, forecast initialization time and location, and were calculated in the first step of the algorithm to find the optimal adjustment period. Once we have established the predicted error in this way, we add it to the EMOS mean forecast $\mu_{t,l}$ and obtain the adjusted RAFT forecast $\hat{\mu}_{t,l}$:
\begin{equation}
\hat{\mu}_{t,l} = \mu_{t,l} + \hat{e}_{t,l}
\end{equation}

\begin{figure*}
  \centering
  \includegraphics[width=0.6\textwidth]{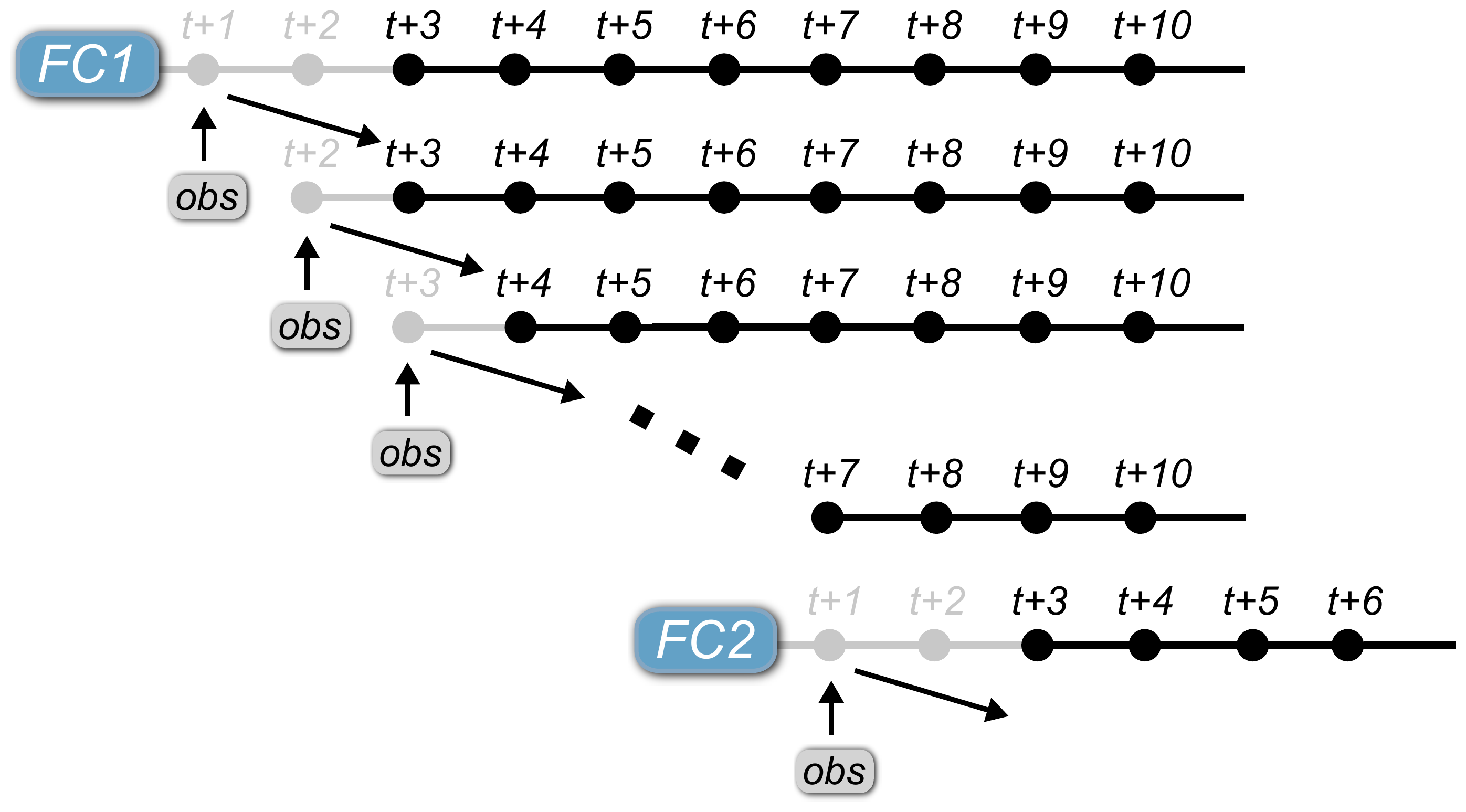}
  \caption{Diagram of a forecast cycle for an hourly forecast issued every six hours with rapid adjustment of the forecast trajectory (RAFT) applied as new observations become available. Forecasts in gray are only used as predictors by means of their observed error and not adjusted themselves.}\label{fig:fig05}
\end{figure*}

The resulting adjusted mean forecast is generated from data that has passed through multiple levels of post-processing. First, while applying EMOS, the performance of the raw ensemble over the past 40 days is analyzed and the results are used to improve the deterministic and probabilistic forecast skill. This post-processing method uses forecasts and observations from a rolling training period and is carried out right after the NWP model run has finished and before the forecast is issued. When the first forecast from the trajectory verifies two hours later, we make the first RAFT adjustment and continue in the same manner in hourly intervals (see Figure~\ref{fig:fig05}). The level of RAFT error correction only relies on the performance of the EMOS forecast mean during the current forecast run, using very short-term information not available when the NWP model was initialized and when EMOS was applied. The combined EMOS/RAFT predictive distribution consisting of the RAFT forecast as mean and the EMOS variance can produce a more accurate forecast than both the raw ensemble and the unadjusted EMOS forecast, while remaining calibrated. 

\section{Results} \label{sec:sec04}

In the previous section, we described how the RAFT method can be combined with post-processing methods like EMOS to provide an additional short-term error correction. We now show comprehensive results, first for Heathrow Airport and then for all sites in the data set. 

\subsection{Results for Heathrow Airport}
As one of the busiest airports in the UK, accurate weather forecast for Heathrow are of major importance, especially for the very short term \citep[e.g.][]{GhirardelliGlahn2010}. Therefore, we investigate the impact of RAFT on forecast quality at this site separately. From Figure~\ref{fig:fig04}, we know how the relationship between forecast errors at different lead times can be used to define the RAFT regression model and corresponding adjustment periods. This analysis is done only once and the parameters are then valid until there are significant changes in the forecast models or the local error characteristics.

\begin{figure*}
  \centering
  \includegraphics[width=0.75\textwidth]{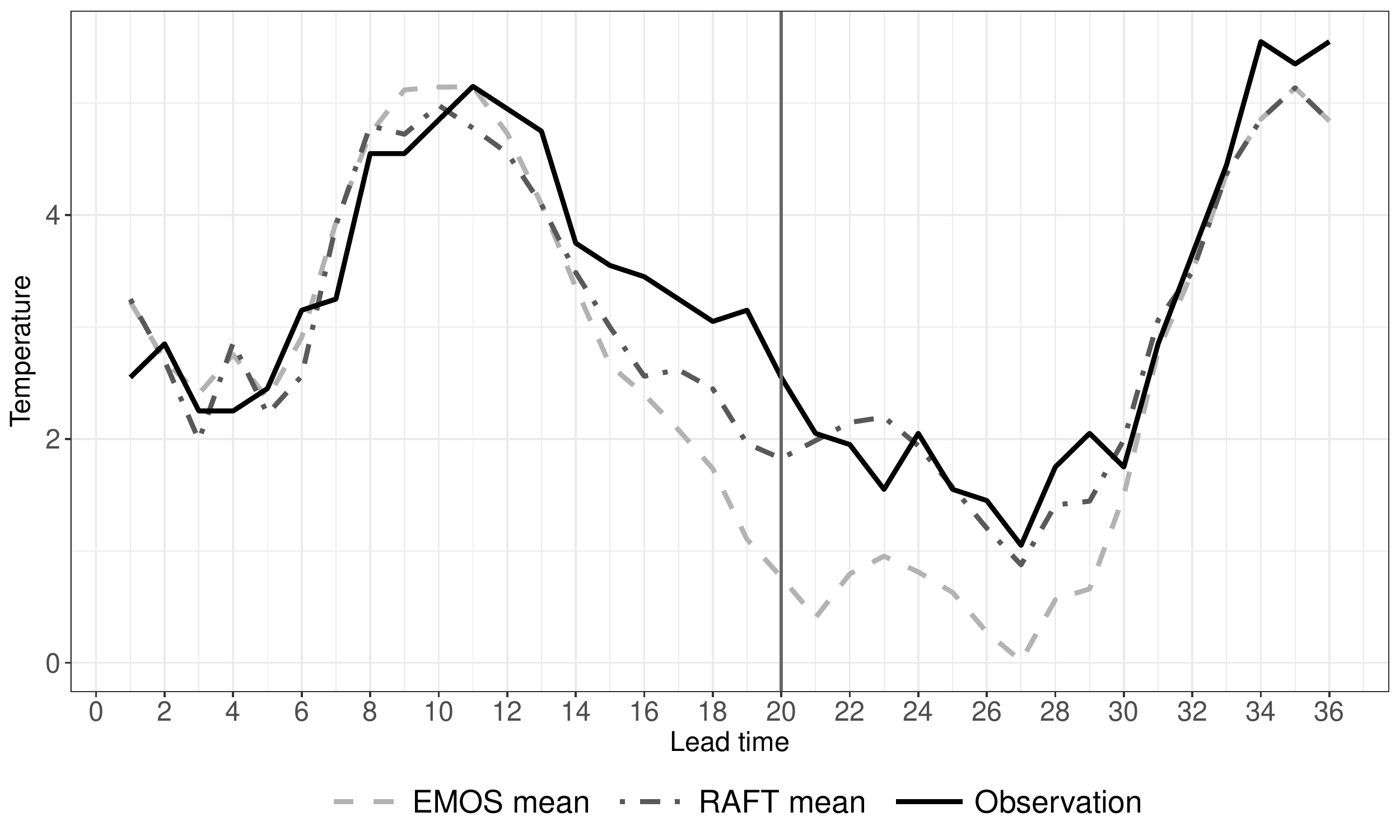}
  \caption{Example forecast at Heathrow Airport. Snapshot of the RAFT and EMOS trajectories taken at January 14, 2016 2300 UTC (corresponding to the vertical line), where the model was initialized 20 hours earlier. See the text for further details.}\label{fig:fig06}
\end{figure*}

In the following example, we illustrate how RAFT works in a real-time setting. Figure~\ref{fig:fig06} is a snapshot, taken at 2300 UTC on January 14, 2016 at Heathrow Airport. The light grey dashed line depicts a forecast trajectory, issued at 0300 UTC the same day and post-processed using EMOS as described in Section \ref{sec:sec2_2}. Over time, temperature values (represented by the black solid line) are observed for the 36 lead times of the trajectory. At the time of the snapshot, they are however only available up to one hour before. The dot-dashed dark grey line is the RAFT forecast and consists of two parts. The trajectory left of the black vertical line is a combination of the most recent RAFT forecasts at each lead time, i.e.\ the forecast issued one hour earlier, using the error information from two hours before the valid time. These are the optimal RAFT forecasts, as they contain the most information and are very short-term. 

\begin{figure*}
 \centering
 \includegraphics[width=0.75\textwidth]{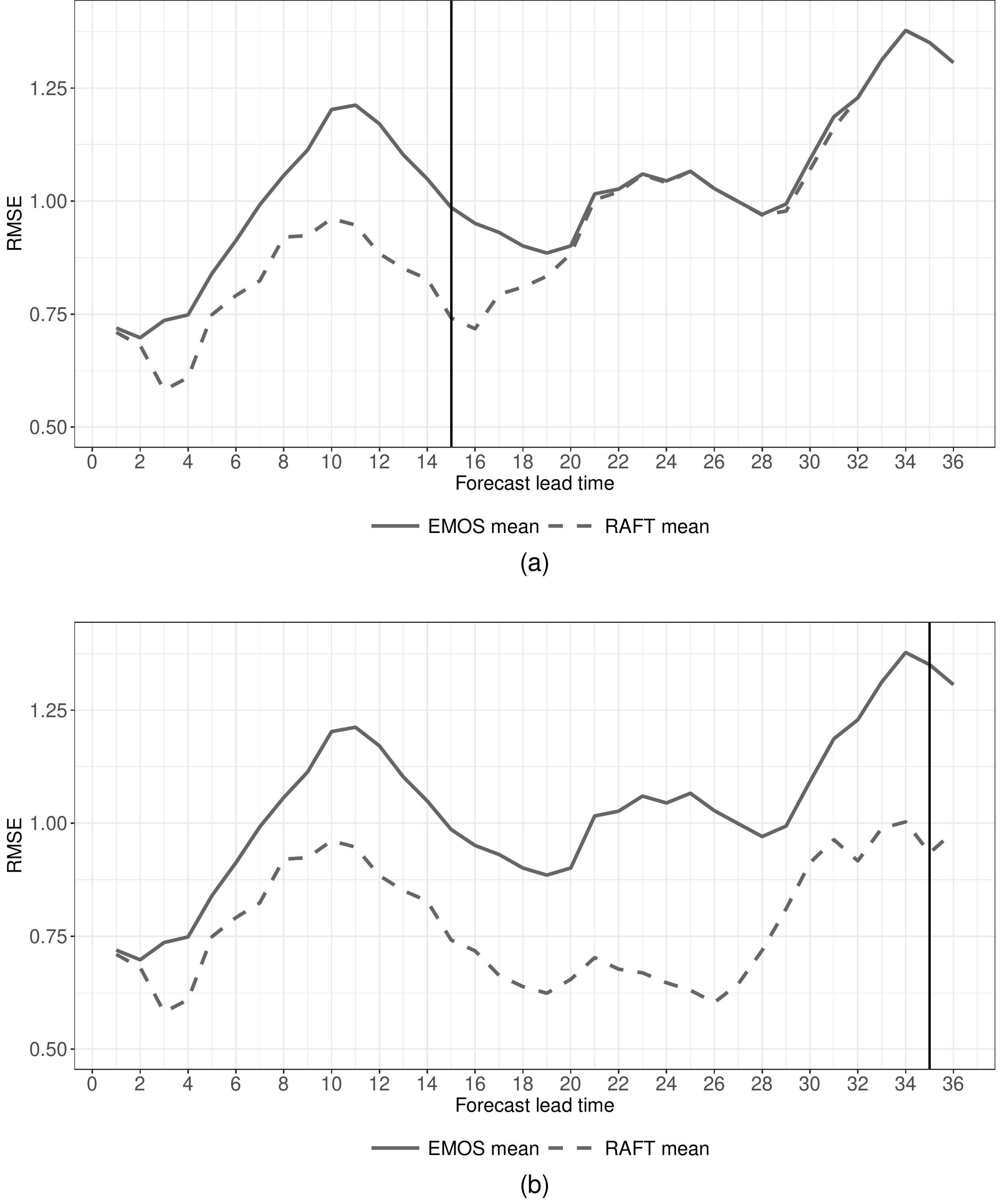}
 \caption{(a) RMSE of the EMOS and RAFT mean forecasts over lead time. The scores are averaged over all dates in the test period at Heathrow Airport for model runs initialized at 0300 UTC. RAFT error corrections are only carried out until lead time $t+15$. (b) Same as (a), but RAFT is carried out for all lead times until the end of the trajectory.}
 \label{fig:fig07}
\end{figure*}

The right side of the black vertical line is the current RAFT trajectory, showing the best possible forecast we can make with the information we have at this point in time. Depending on the length of the adjustment period, the forecasts from here to the end of the original forecast trajectory are adjusted using the most recent error information. For example, the forecast at $t+28$ is being adjusted, while the forecast at $t+33$ is not. For the first twelve hours, the uncorrected trajectory has a good agreement with the observations and only small corrections are made. Between lead times 15 and 30, corresponding to evening and night time, the EMOS forecast underpredicts the temperature. As soon as larger errors are observed, the RAFT adjustment to the original forecast also becomes larger and after a short time manages to counter the underprediction. This example illustrates how RAFT is able to quickly correct forecast errors a few hours ahead, whereas the unadjusted forecast would continue to underpredict the temperature for further 15 hours.

To evaluate the performance of RAFT over the entire test period, we look at the root-mean-square error of the RAFT-adjusted forecasts and compare to the unadjusted EMOS mean forecasts. Figure~\ref{fig:fig07} shows the RMSE at Heathrow, averaged over all cases in the test period where the NWP model was initialized at 0300 UTC. In both plots, the solid line is identical and represents the performance of the EMOS-post-processed forecast trajectories, and the dashed line is the RMSE of the RAFT forecasts. The difference between the plots lies therein that they are snapshots taken at different points in the forecast cycle. 

The upper plot (Figure~\ref{fig:fig07}a) depicts the level of forecast skill if we stopped applying RAFT after lead time $t+15$. This would mean that all forecasts to the left of the vertical line have been adjusted according to the forecast error measured two hours earlier. As the most recent observed error is registered at $t+14$, all forecasts to the right of the vertical line are adjusted using this error information (depending on the length of the respective adjustment periods). This means that on the left side, the difference between the two curves is the maximum improvement obtainable by applying RAFT.

For the first few hours, there is only very little improvement, as we don't have any information about the current run's forecast error yet and we have to rely on the information from the run started 24 hours earlier. However, as soon as the new error information is available, RAFT shows a considerable reduction in forecast error, even up to 20\%. On the right side, the largest benefit can be seen in the next few hours, as the correlation is strongest between close lead times. After about five hours, RAFT falls back to the skill level of the EMOS forecasts. Interestingly, for the period between $t+28$ and $t+32$, there appears to be a significant correlation to the error at $t+14$. Thus we see a small error reduction 14 to 18 hours ahead.

In Figure~\ref{fig:fig07}b, a different snapshot is shown. Now we apply RAFT to the full forecast cycle, i.e. let it run until the last adjustment made at $t+35$. This plot considers only the most short-term correction for each lead time and therefore the best possible forecast. Here we see a large improvement over EMOS throughout and especially for later lead times. An interesting feature emerges if we compare the forecast skill at lead times $t+2$ and $t+26$. These lead times correspond to the same time of day, 0500 UTC, and we would expect the forecasts at $t+26$ to perform worse due to more time having passed since the model initialization. With RAFT, however, this forecast was adjusted with a very recently measured observation error, whereas the $t+2$ forecast could only be adjusted using the data from the model run initialized 24 hours prior. As a result, the $t+26$ error is lower than the one at $t+2$ and, consequently, a forecast for $t+26$ of an older model run will on average have more forecast skill than the $t+2$ forecast from the next (and newer) model run.

\begin{figure*}
  \centering
  \includegraphics[width=\textwidth]{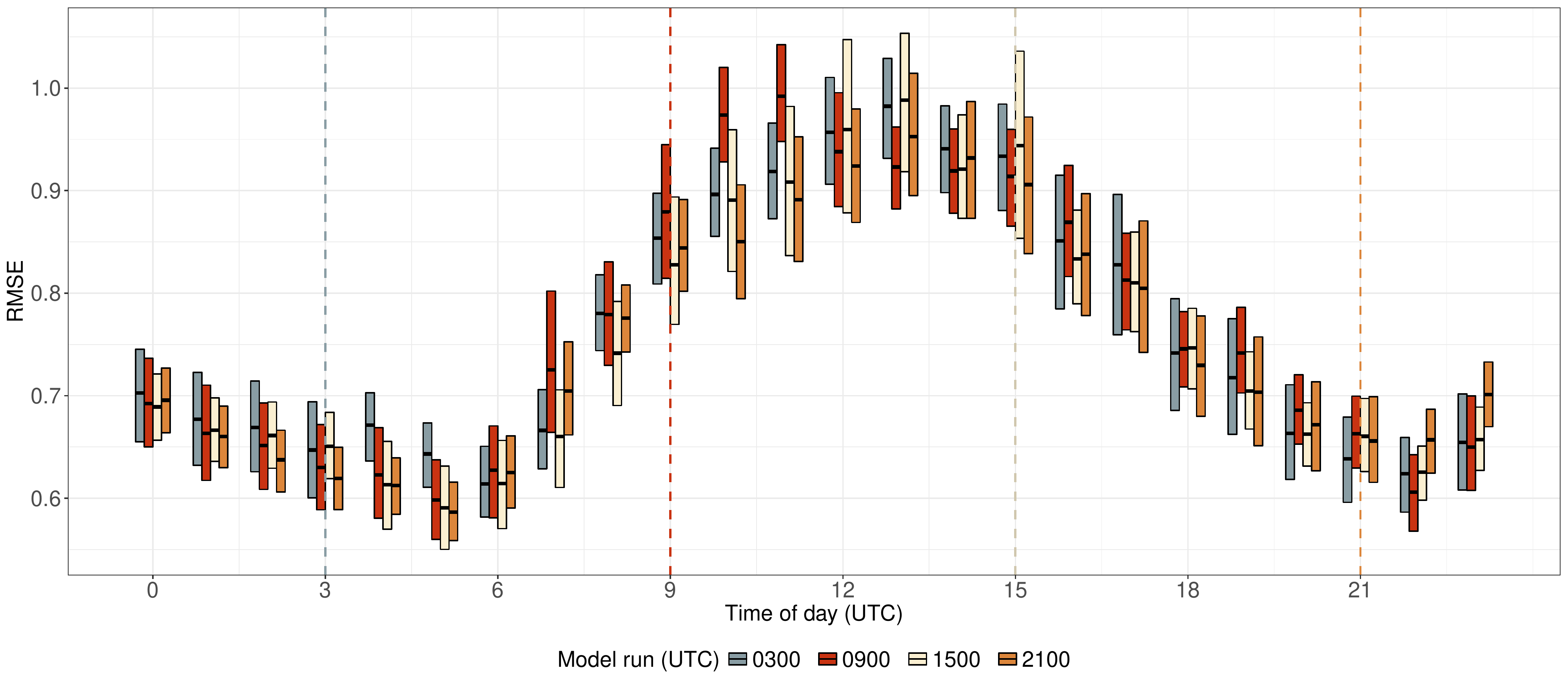}
  \caption{Average root-mean-square error for all four daily NWP model runs as a function of the time of day. The scores were computed over the test period at Heathrow Airport and are shown with 90\% bootstrap confidence intervals. The dashed vertical lines represent the initialization times of the NWP model.}\label{fig:fig08}
\end{figure*}

This means that there is a transition period at the beginning of every NWP model run, where an old run provides better forecasts until the point is reached where the forecasts from the new model run can be used for the RAFT adjustment. Figure~\ref{fig:fig08} illustrates the relationship between all four initialization times, depicting the average RAFT RMSE as a function of the time of day in UTC. The times when a new model run is started are marked by dashed vertical lines. Again, the RMSE is computed using the most recent adjusted and optimal forecast. Here, the mean score is shown, as well as 90\% confidence intervals based on 1000 bootstrap samples. 

At first glance, there is a strong diurnal variation in all four runs, with the lowest predictability around midday and the highest during the early morning. We are interested in the ranking of the four runs in terms of forecast skill. Ordinarily, we would expect the newest run to be the best, but as seen in Figure~\ref{fig:fig07}b, there is a short period during which an older run produces better forecasts. For the first few hours of the day, the ranking is as expected, in that the 2100 UTC run has the lowest RMSE and the 0300 UTC run the highest. When the first forecast from the new 0300 UTC run comes in at 0400 UTC, the skill decreases considerably, instead of improving. This is due to the fact that there is no recent forecast data available for the RAFT adjustment and we have to rely on the error information from 24 hours before. For two hours after the initialization of the 0300 UTC run, the 2100 UTC run remains the best forecast; the score difference between the two runs is actually significant at the 90\% level. After 0600 UTC, the model runs rank in the expected order. 

A similar pattern can be noticed every time a new model run is produced, with the exception being the 1500 UTC run. This run actually ranks best, or at least close to the others, from the first forecast, coinciding with the increase in predictability in the afternoon. We can conclude that the four daily model runs have comparable forecast quality after applying RAFT, apart from a transition period of about two hours. During this period, forecasts from an older run should be preferred to the newest.

\subsection{Results for all sites}

\begin{figure*}
 \centering
 \includegraphics[width=0.75\textwidth]{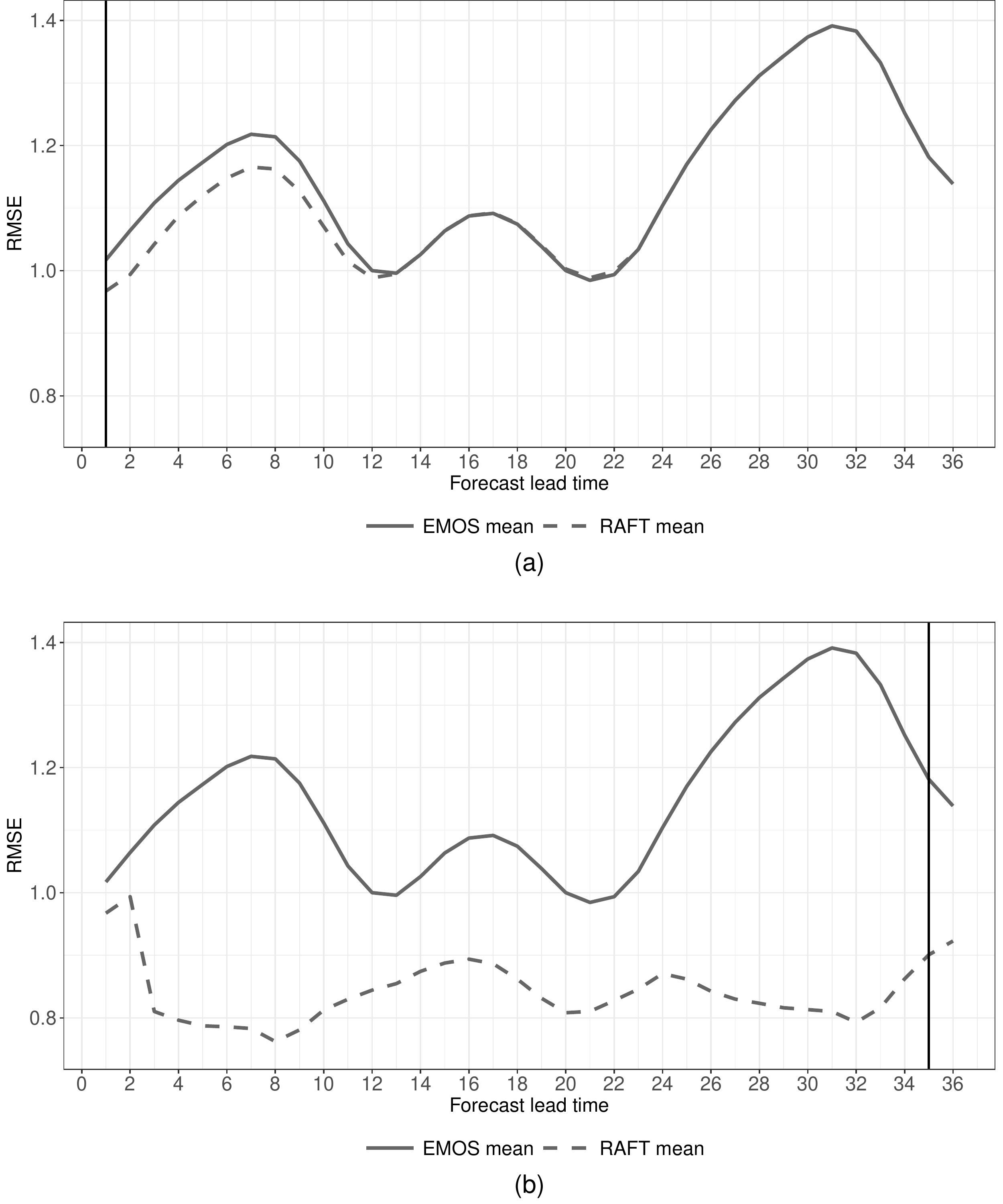}
 \caption{(a) RMSE of the EMOS and RAFT mean forecasts over lead time. The scores are averaged over all dates and locations in the test period for model runs initialized at 2100 UTC. RAFT error corrections are only carried out once at lead time $t+1$. (b) Same as (a), but RAFT is carried out for all lead times until the end of the trajectory.} \label{fig:fig09}
\end{figure*}

After presenting the results for Heathrow Airport, we now discuss how RAFT performs for all observation sites available. The dataset covers the British Isles (Figure~\ref{fig:fig02}) and displays a wide variety of local characteristics, like sites in the Scottish mountains at elevations above \unit{1000}{m} or coastal towns.

Figures~\ref{fig:fig09}a and \ref{fig:fig09}b compare the average RMSE of the EMOS and RAFT forecasts for the 2100 UTC model run, similar to Figure \ref{fig:fig07}. Again, they represent snapshots at different times in the RAFT adjustment process. In Figure~\ref{fig:fig09}a, we see the maximum achievable RAFT improvement over the EMOS mean if we only applied the adjustment once at the moment the first forecast becomes valid at $t+1$. At that time, no observations are available yet for the new run, so we have to rely solely on error information from the run initialized 24 hours earlier. Those RAFT forecasts for which the adjustment period extends beyond the beginning of the run have been adjusted using the observation made at $t+0$, combined with the old run's $t+24$ forecast.

While the benefit from applying RAFT in this way is considerably smaller than the improvement we see as soon as the new forecast data is used, there is still a reduction in the RMSE for the next twelve hours. We notice an interesting detail between $t+20$ and $t+23$ (corresponding to 1700 UTC and 2000 UTC, respectively). In this period of high predictability, the RAFT scores are actually slightly worse than the EMOS scores, but revert to being equal with the next RAFT adjustment at $t+2$ (not shown). This pattern can be observed at a handful of sites, where the error correlation between the lead times is particularly strong and the corresponding adjustment periods quite long. The RAFT algorithm for determining the adjustment period does not take into account any potential stark differences in correlation patterns between the sites. It might therefore be advisable to look into adjusting the algorithm if interested in optimizing the performance for specific locations. 

In Figure~\ref{fig:fig09}b, we again see the outcome if RAFT is applied every hour up until the last installment at $t+35$. This represents the maximum and most short-term gain in forecast skill achievable at every lead time and is not a continuous trajectory. We will use these forecasts for the entire subsequent analysis. At the beginning of the forecast cycle, there is a sharp drop in the RMSE, immediately after we are able to use data from the current run. Afterwards, the RAFT skill remains relatively constant, with small variations due to the diurnal cycle, whereas the EMOS skill fluctuates considerably. Especially during the last twelve hours of the forecast cycle, the improvement of RAFT over EMOS is quite substantial, as the short-term RAFT forecast corrections manage to cancel out the skill deterioration usually occurring with increasing lead time.

\begin{figure*}
  \centering
  \includegraphics[width=\textwidth]{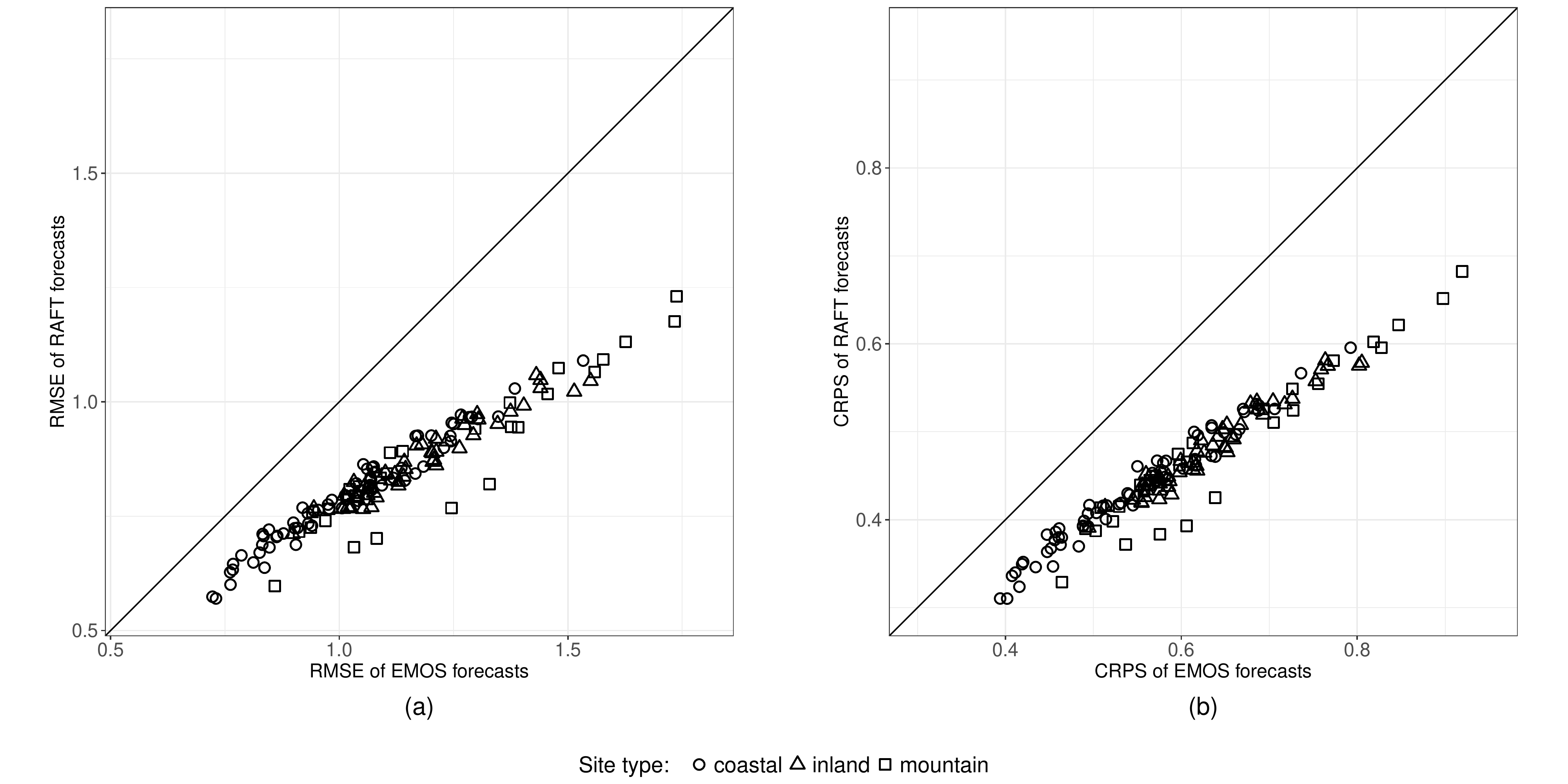}
  \caption{(a) Average RMSE of the RAFT forecasts as function of the EMOS RMSE for all sites, lead times and model runs in the data set. (b) Average CRPS of the RAFT forecasts as function of the EMOS CRPS for all sites, lead times and model runs in the data set. The RAFT CRPS is computed using the EMOS predictive variance.}
  \label{fig:fig10}
\end{figure*}

All observation sites in the study can be separated into three categories based on their location: coastal, inland and mountain sites. In Figure~\ref{fig:fig10}, the RMSE and CRPS scores for all locations are aggregated over all four model runs and the RAFT scores are plotted against the EMOS scores. The CRPS for RAFT is calculated by plugging the RAFT mean into the EMOS predictive distribution. For both the RMSE and the CRPS, we see an improvement for all sites after applying RAFT, in particular at locations where the error was high in the first place. In fact, the improvements seem to follow the same linear trend, apart from a group of five mountain sites (located in Scotland and Cumbria), which receive a somewhat larger benefit from RAFT than the other sites. This hints at some location-specific issues not resolved by EMOS or the original ensemble.

\begin{figure*}
  \centering
  \includegraphics[width=0.85\textwidth]{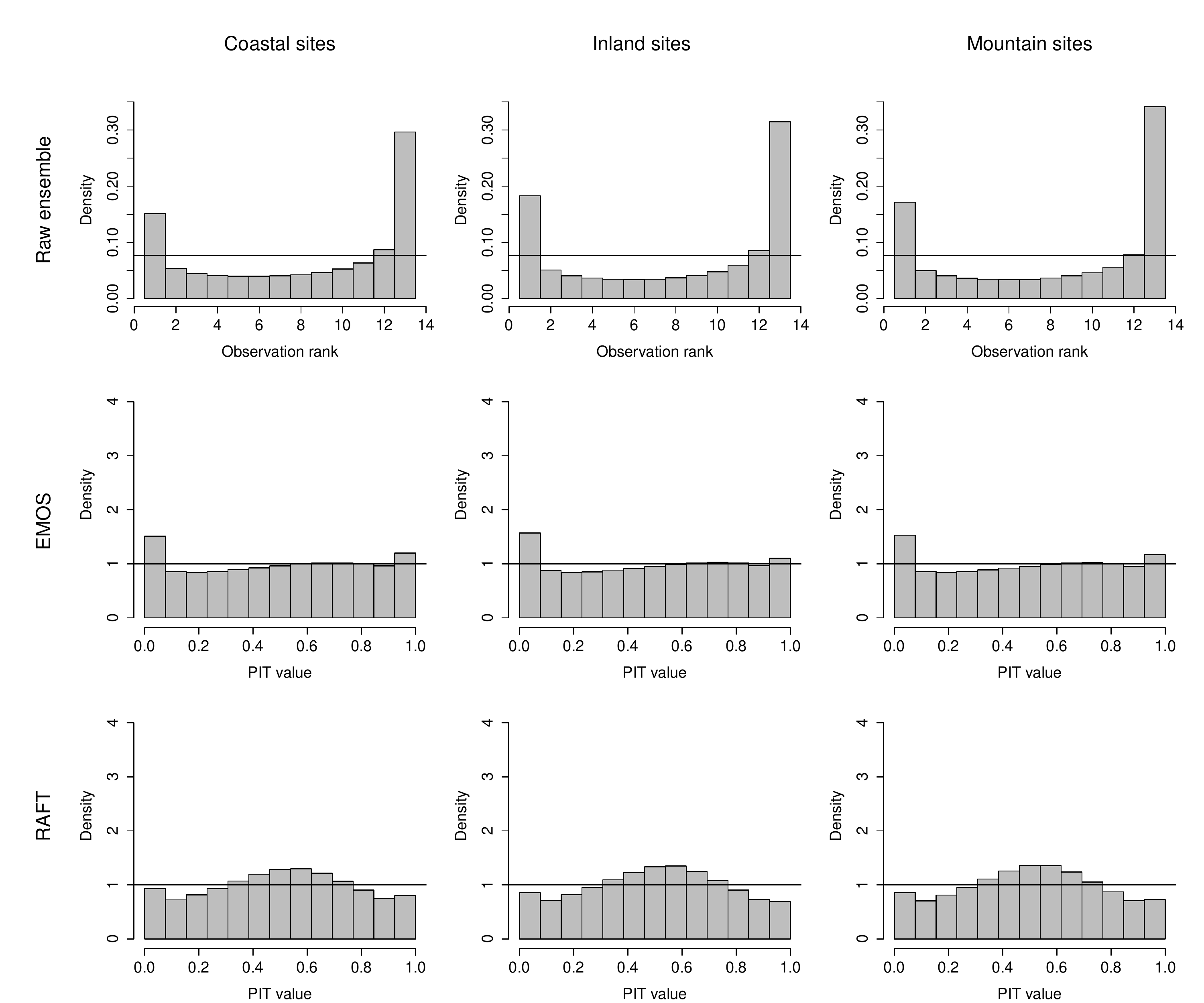}
  \caption{Verification rank histograms for the raw ensemble (top row) and PIT histograms for the EMOS (middle row) and RAFT (bottom row) forecasts. The RAFT predictive distribution is generated by using the EMOS predictive variance. The histograms are divided by site type and data is aggregated over all dates and model runs in the test set.}\label{fig:fig11}
\end{figure*}
 
In Figure~\ref{fig:fig03}, we showed that EMOS produces nearly calibrated forecasts and naturally we want to preserve this level of calibration with RAFT. Therefore we compare the rank and PIT histograms of the raw ensemble, EMOS and the distribution consisting of the RAFT mean and the EMOS predictive variance. Figure~\ref{fig:fig11} shows these histograms divided by site type. For all three forecasting methods, there is only very little difference in calibration between coastal, inland and mountain sites. The raw ensemble is, as expected, uncalibrated and very underdispersive, recognizable by the characteristic U-shape. EMOS is fairly calibrated, although there is still some hint of a bias and underdispersion. In contrast, RAFT is slightly overdispersive, meaning that the variance of the distribution is on average too large. This is not surprising, given that the mean of the distribution now has much better deterministic skill, but the corresponding EMOS variance has not changed. An additional adjustment of the EMOS variance to counteract the induced overdispersion is a potential subject for further study.

Another indicator of calibration is the actual coverage of the prediction interval compared to the nominal value. The ensemble members create a prediction interval of $11/13 \approx 84.62\%$, which would correspond to perfect calibration. However, the raw MOGREPS-UK ensemble only reaches a coverage of $52.24\%$, whereas the EMOS coverage is $79.29\%$ and the RAFT prediction intervals cover $87.31\%$. Although one is under- and the other overdispersive, both EMOS and RAFT are nearly calibrated, with RAFT being actually slightly closer to the correct value. 

\begin{figure*}
  \centering
  \includegraphics[width=0.8\textwidth]{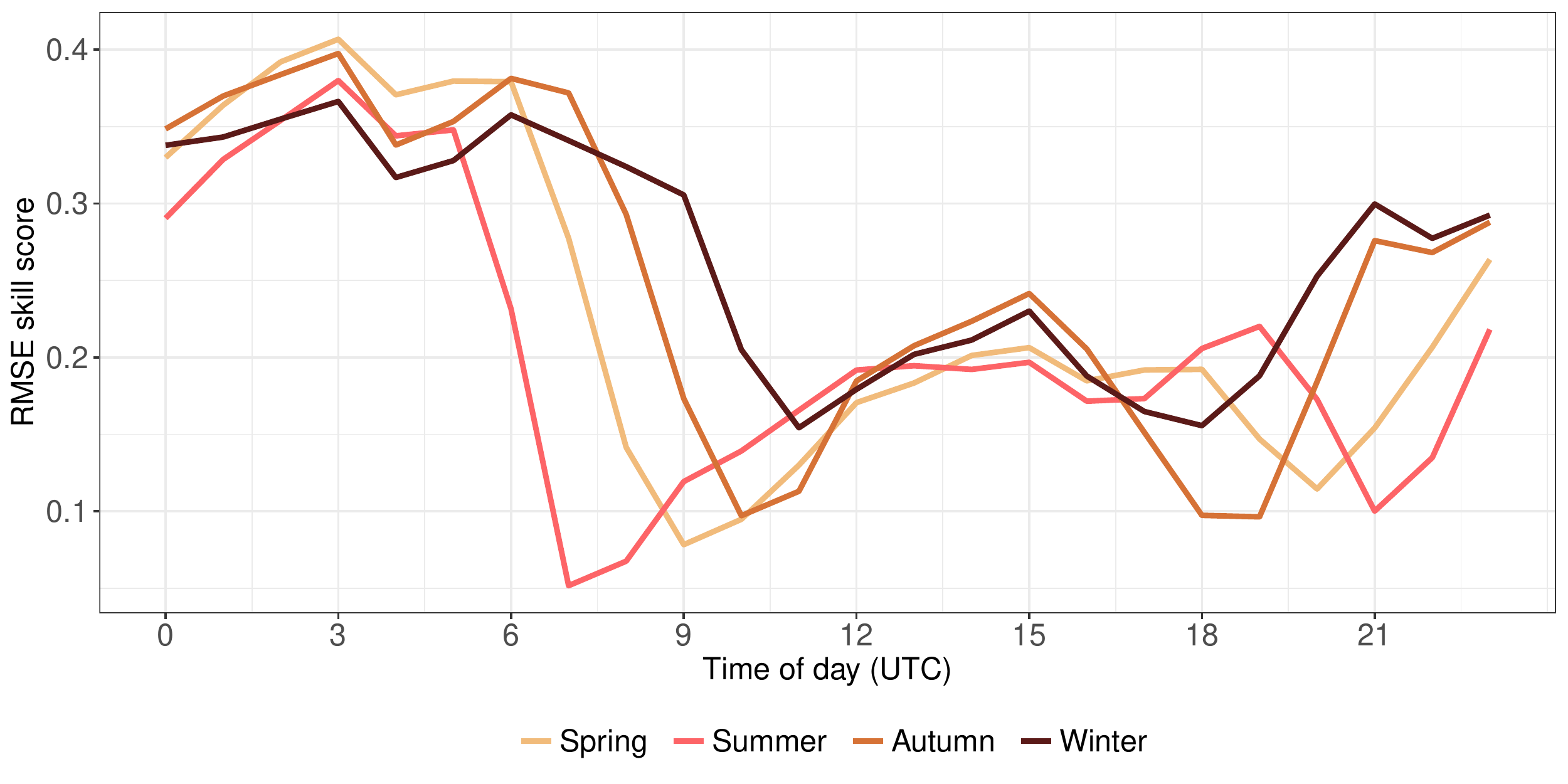}
  \caption{RMSE skill score of RAFT with EMOS as reference forecast against time of day for different seasons. RMSE scores are averaged over all sites, model runs and dates in the test data set.}\label{fig:fig12}
\end{figure*}

Finally, we look at how RAFT performs during different seasons of the year. The test set contains two full spring seasons, and one full winter, summer and autumn. Figure~\ref{fig:fig12} depicts the RMSE skill score, the relative improvement of the RAFT over the EMOS mean, for the four seasons. A score of 1 would mean a perfect forecast and a score of 0 no improvement over the reference forecast. Again, all four runs and all sites have been aggregated. 

The largest gain in forecast skill occurs during the night and is very similar for all seasons. The same pattern holds for the time between 1200 UTC and 1600 UTC, where the skill score values are very close. In the morning, however, the scores for summer and winter behave very differently; they both decrease, but the summer skill score much faster and further than the winter score. This is due to the fact that in summer, the diurnal cycle plays a much more prominent role (not shown) and the predictability during night is much higher than during the day. In winter, the RMSE is more stable and there is only very little difference in predictability. The deterioration in the skill score during the early morning in summer coincides with a period of large change in predictability. It seems that during this time predictability changes so fast that even the very short-term RAFT adjustment can only improve the forecast skill by a small amount. Therefore, it might be advantageous to look into obtaining separate RAFT coefficients for the different seasons. This is not possible in the context of the current study, however, as a much larger training data set would be required. 

\section{Conclusions}\label{sec:sec05}
This paper presents a new post-processing approach for NWP forecasts, rapid adjustment of forecast trajectories (RAFT), that is applied on top of the traditional post-processing approach EMOS once new information pertaining to the current forecast trajectory becomes available. By utilizing the forecast error correlation structure in the post-processed NWP forecast trajectories, the EMOS mean forecasts of the not-yet-realized part of the trajectory are adjusted in every time step of the forecast based on the forecast errors that have already been realized. This computationally efficient approach to make use of the newest available information provides an appealing alternative to computationally costly rapid ensemble cycles \citep{Lu&2007, Benjamin&2016}, and the older forecast gains skill in the time between initialization and release of the next NWP forecast cycle.

While the precise setup described here may have some operational restrictions due to computing and observation processing time if applied at a large number of locations, our results provide a convincing proof-of-concept. For example, as shown in Figure~\ref{fig:fig09}b, the forecast skill may be improved by over 40\% on average in terms of RMSE when a 32-hour-old forecast is supplemented with the most recent available information an hour before it is realized. Our results at Heathrow Airport furthermore suggest a new strategy for updating the forecast cycle in that a delay in introducing the new NWP forecast may be preferred if RAFT is employed.    

RAFT is easily implemented at individual locations and could be especially useful for applications such as aviation and renewable energy production where decision-making relies on location-specific skillful weather forecasts. In such cases, observation frequency may be higher than the time resolution of the NWP forecast, a situation to which RAFT can easily be adapted. In our analysis, we only update the mean of the EMOS forecasts while the variance remains unchanged. The original EMOS forecasts are slightly underdispersive and biased; a similar effect has been reported in previous applications of EMOS to individual locations, see e.g. \citet{ThorarinsdottirGneiting2010}. The RAFT procedure reduces the bias and improves the overall calibration, while changing the sign of the miscalibration to slightly overdispersive, cf. Figure~\ref{fig:fig11}. Our experiments to update the EMOS variance simultaneously with the mean were not successful in that they did not result in further skill improvement. However, this might be worth investigating further in cases where the originally post-processed forecast is nearly perfectly calibrated, or slightly overdispersive. 

\section*{Acknowledgements}
We thank the Met Office, in particular the IMPROVER team, for giving us the possibility to work with MOGREPS-UK data. Nina Schuhen acknowledges the support of the Research Council of Norway through grant nr. 259864 ``Stipendiatstillinger til Norsk Regnesentral''. Alex Lenkoski acknowledges the support of the Research Council of Norway through grant nr. 237718 ``BIG INSIGHT -- Statistics for the knowledge economy''.

\bibliographystyle{wileyqj}

\end{document}